\documentclass[]{spie}  

\newcommand{\prox}{{\mathrm{prox}}}

\usepackage{amsmath,amsfonts,amssymb}
\usepackage{tikz}
\usetikzlibrary{spy}
\usepackage{graphicx}
\usepackage{array}
\usepackage[colorlinks=true, allcolors=blue]{hyperref}
\usepackage{algorithm}%
\usepackage{algorithmicx}%
\usepackage{algpseudocode}%
\usepackage{listings}%
\usepackage[percent]{overpic}
\usepackage{xcolor}
\usepackage{pgfplots}
\usetikzlibrary{spy,calc}
\usepackage{multicol}
\usepackage{subcaption}
\usepackage{multirow} 

\title{Sparse Measurement Medical CT Reconstruction using Multi-Fused Block Matching Denoising Priors}

\author[a]{Maliha Hossain}
\author[b]{Yuankai Huo}
\author[b]{Xinqiang Yan}
\author[a]{Xiao Wang}
\affil[a]{Oak Ridge National Laboratory, 1 Bethel Valley Road, Oak Ridge, TN, USA 37830}
\affil[b]{Vanderbilt University, 2201 West End Ave., Nashville, TN, USA 37235}

\authorinfo{Further author information: (Send correspondence to Maliha Hossain.)\\Maliha Hossain.: E-mail: hossainm@ornl.gov}

\begin{document} 
\maketitle

\begin{abstract}

A major challenge for medical X-ray CT imaging is reducing the number of X-ray projections to lower radiation dosage and reduce scan times without compromising image quality.
However these under-determined inverse imaging problems rely on the formulation of an expressive prior model to constrain the solution space while remaining computationally tractable. Traditional analytical reconstruction methods like Filtered Back Projection (FBP) often fail with sparse measurements, producing artifacts due to their reliance on the Shannon-Nyquist Sampling Theorem. 
Consensus Equilibrium, which is a generalization of Plug and Play, is a recent advancement in Model-Based Iterative Reconstruction (MBIR), has facilitated the use
of multiple denoisers are prior models in an optimization
free framework to capture complex, non-linear prior information. However, 3D prior modelling in a Plug and Play approach for volumetric image reconstruction requires long processing time due to high computing requirement.

Instead of directly using a 3D prior, this work proposes a BM3D Multi Slice Fusion (BM3D-MSF) prior that uses multiple 2D image denoisers fused to act as a fully 3D prior model in Plug and Play reconstruction approach. 
Our approach does not require training and are thus able to circumvent ethical issues related with patient training data and are readily deployable in varying noise and measurement sparsity levels. 
In addition, reconstruction with the BM3D-MSF prior achieves similar reconstruction image quality as fully 3D image priors, but with significantly reduced computational complexity.
We test our method on clinical CT data and demonstrate that our approach improves reconstructed image quality.
\end{abstract}


\section{INTRODUCTION}
\label{sec:intro}  

X-ray Computed Tomography (CT) is a widely used medical imaging technique that provides non-invasive visualization of the interior radiodensity distribution of the patient, aiding in screening and diagnosis. The CT system consists of a rotating gantry that revolves around the patient. One end of the gantry is an X-ray source and the other end is an X-ray detector, capturing a series of X-ray projection measurements at different view angles of rotations.
These view measurements at different angles are repeated for cross-sections along the transverse plane and then reconstructed into a volumetric image of the patient.

A significant area of research in medical CT is reducing the number of X-ray projections. This can be achieved either by decreasing the number of projection view angles per rotation, reducing the acquisition pitch, or a combination of both. 
Such sparse measurement acquisition presents many benefits, including shorter scan times \cite{kudo2013image, wang_2021}, lower radiation dose~\cite{Qu20,Zhao18, wang_2021}, and reduced facility costs due to faster CT scan turnaround times~\cite{wang_2021}. 
Notably, patients' taken radiation doses increase approximately linearly with the number of X-ray projections. 
Thus, by reducing the number of X-ray projections and radiation doses, the radiation-induced cancer risk for patients can be much lower~\cite{kudo2013image,zheng2022low}.
However, producing high-quality reconstructions from sparse measurements acquired at sub-Nyquist rates presents significant image reconstruction challenges. 
Analytical methods, such as Filtered Back Projection (FBP), are commonly used in clinical practice. 
They are based on the Radon Transform \cite{toft1996radon} and adhere to Shannon-Nyquist Sampling Theorem. 
Consequently, they are often unable to accurately reconstruct images from sparse projections sampled below Nyquist rates, producing streaking artifacts that are evident in the FBP reconstructions shown in figures \ref{fig:brain_recon} and \ref{fig:brain_recon}.
as well as in literature~\cite{wang_2021,hossain2020ultra, thibault2007three}.

Established methods for sparse measurement CT reconstruction fall into three main categories: measurement domain preprocessing~\cite{9263571,cheng2023sparse}, image domain postprocessing~\cite{han2016sparse,zhang2018sparse}, and statistical iterative reconstruction \cite{zhang2018sparse,wang_2021,thibault2007three}.
The preprocessing approach involves missing measurements augmentation, either using interpolation~\cite{9263571}
or neural network based methods~\cite{cheng2023sparse}, followed by image reconstruction.
In postprocessing, an initial image reconstruction is further enhanced by removing artifacts by using sparse transforms 
\cite{han2016sparse} and neural network \cite{zhang2018sparse} based methods.
However both of these approaches are limited in that they may produce hallucinated image features due to the pre-processing or post-processing results not constrained by CT acquisition physics.
Statistical Iterative Reconstruction approaches, such as Model-Based Iterative Reconstruction (MBIR) \cite{wang_2021, thibault2007three, hsieh2013recent}, offer a more robust solution by formulating a Maximun a Posteriori estimation problem. MBIR employs Bayesian estimation and linear algebra to construct a statistical image prior model and a data fidelity forward model. The image prior model represents the statistical appearance of the reconstructed image, while the data fidelity forward model aligns the reconstructed image with the X-ray projections and CT geometry. By optimizing a cost function that integrates both models, MBIR can accurately reflect the acquisition physics and scanner geometry, producing clearer images with fewer artifacts, especially at low radiation doses and limited projections. The MBIR approach won the 2022 AAPM Grand Challenge for low-dose CT image reconstruction, demonstrating its superior image quality~\cite{abadiaapm}.

Despite its success, MBIR's reliance on statistical priors such as Markov Random Fields (MRF) \cite{thibault2007three} that must fit within a cost function formulation limits their ability to model complex image features in sparse measurement reconstruction. 
The result is overly smoothed reconstructions as in 
figure \ref{fig:brain_recon}(c).
Recent developments in MBIR such as 
Plug and Play (PNP) and its generalization, Consensus Equilibrium \cite{buzzard2018plug} have shown that the use of deep learning image denoiser, such as denoising Convolutional Neural Networks, as image prior models can represent image features more faithfully than traditional statistical image priors, achieving high-quality reconstructions \cite{majee2021multi,venkatakrishnan2021algorithm}.
However, neural networks depend heavily on training images, potentially leading to misleading findings for patients with outlier features that are not represented in the training dataset. 
In contrast, block matching algorithms for 2D images, BM3D,  \cite{dabov2006image, maggioni2012nonlocal} 
and its 3D variant, BM4D, ~\cite{Maggioni13}
remain state-of-the-art among off-the-shelf denoisers that do not need any training. 
By avoiding pretraining, these denoisers avoid the risk of hallucination or out-of-distribution artifacts, which could lead to misdiagnosis.
The fusion of multiple lower dimensional denoisers, in particular convolutional neural networks, to form a higher dimensional prior model for inverse imaging in a multi agent consensus equilibrium framework was coined as Multi-Slice Fusion (MSF) in \cite{majee2021multi}. 
The use of block matching denoisers with MSF was demonstrated in \cite{hossain2020ultra} for inverting ultra-sparse CT in industrial inspection. 
\cite{sridhar2020distributed} presented strategies to exploit the parallelism inherent in a multi agent consensus equilibrium routine to speed up performance.
However, this prior modelling approach has only been tested for material science phantoms and has not been established for medical CT imaging on real patient data. 
In addition, neural network based denoising priors require significant amount of computing resource for training, making them infeasible for clinical practice~\cite{hossain2020ultra}.

In this paper, we propose a BM3D-Multi Slice Fusion (BM3D-MSF) reconstruction framework that leverages off-the-shelf block matching denoisers for medical CT reconstruction, eliminating the need for training while surpassing the image quality of conventional MBIR. Our novel fusion technique significantly reduces the computational complexity associated with block matching denoiser priors. Rather than performing image prior operations for the entire 3D image volume, as done in the BM4D algorithm, we utilize three 2D priors corresponding to three orthogonal spatial planes ($X-Y$, $Y-Z$, and $X-Z$). These 2D prior results are then fused together, effectively reducing the computational demands of a 3D prior to a series of 2D priors without compromising image quality. This prior fusion approach also offers the flexibility to independently adjust the regularization based on the spatial resolution and measurement density along each directional axis.

\section{METHODS}
In this section we outline our methodology for formulating a 
PNP approach for solving inverse imaging problems with a BM3D-MSF prior model and a parallel beam CT forward model using proximal operators. 
The proximal operator of a function $s$ evaluated at $x$ is defined as  
$ \prox_{s} (x) = \arg \min_z \left\{ s(z) + \frac{1}{2}\left\lVert z - x \right\rVert_2^2 \right\}$.
Consider the maximum a posteriori cost function in \ref{eq:map_gen1}
\begin{equation}
\hat{x} = \arg\min_{x}\left\{\sum_{l=1}^{4}s_i(x)\right\} ,
\label{eq:map_gen1}
\end{equation}
where $s_1(x) := \frac{1}{2} \left\lVert y - Ax \right\rVert_{\Lambda}^2$ 
is the data fidelity operator with observation vector $y$, and scanner matrix $A$. $\Lambda$ is the estimated inverse covariance of the noise model and $x$ is the latent image vector of density values. 
$s_2$ through $s_4$ are 2D prior models enforcing regularity assumptions 
along orthogonal spatial planes. 
We reformulate this problem to a constrained one that can then be solved 
using the augmented Lagrangian:
\begin{equation}
    \hat{x},\hat{z} = \arg\min_{x=z}\left\{\sum_{l=1}^{4}s_l(z_l)\right\} = \arg\min_{x,z}\left\{\sum_{l=1}^{4}s_l(z_l)
    +\frac{\rho}{2}\left\Vert x - z_l + u \right\Vert_2^2\right\},
    \label{eq:aug_lgrng}
\end{equation}
where $u = \left[u_1^T,\dots,u_L\right]^T$, $u_l\in \mathbb R^{M_l}$
is the scaled dual variable.
This reformulation allows us to optimize the augmented Lagrangian with respect to $x$ and $z_i$ independently and is the basis for the Alternating Direction Method of Multipliers \cite{boyd2011distributed}. 
By collecting like terms and simplifying, we obtain the following iterative updates:
\begin{align}
&\hat{x}^{(k+1)} = \frac{1}{2}\sum_{i=1}^{L} \left( \hat{z}^{(k)}_l - u^{(k)}_l \right) ,  
    \label{eq:pnp_x}\\
&\hat{z}_l^{(k+1)} = \arg\min_{z} \left\{ s_l(z_l) + \frac{\rho}{2} \left\lVert z_l - u_l^{(k)} - x^{(k+1)}  \right\rVert_2^2  \right\}, 
    \label{eq:pnp_z} \\
&u_l^{(k+1)} = u^{(k)}_l + \hat{x}^{(k+1)} - \hat{z}^{(k+1)}_l 
    \label{eq:pnp_u}.
\end{align}
$\hat{z}_1$ is the solution to the proximal operator of the data fidelity term. 
The proximal operators for statistical priors, as is the case for $\hat{z}_2$ through $\hat{z}_4$, 
have the interpretation of the maximum a posteriori estimate of a  Gaussian denoising operation \cite{meinhardt2017learning} on its noisy argument $\hat{x}^{(k+1)}+u_l^k, l=2,3,4$.
The premise of PNP is that we can substitute these operation with advanced denoisers such as Block Matching algorithms that do not have an explicit cost function form. 
So $s_2(\hat{x}^{(k+1)}+u_2^k)$ is replaced with a BM3D denoiser operating along the $x-y$ plane with denoising strength $\sigma_{x-y}$. 
This operator is denoted $BM3D_{x-y}\left((u_2^{(k)}+\hat{x}^{(k+1});\sigma_{x-y} \right)$. Likewise, we replace $s_3(\hat{x}^{(k+1)}+u_3^k)$ and $s_4(\hat{x}^{(k+1)}+u_4^k)$ with $BM3D_{y-z}\left((u_3^{(k)}+\hat{x}^{(k+1});\sigma_{y-z} \right)$ and $BM3D_{x-z}\left((u_4^{(k)}+\hat{x}^{(k+1});\sigma_{x-z} \right).$
This fusion of multiple lower dimensional denoisers to form the 3D prior model with tunable regularization along different spatial axes produces out proposed prior model denoted BM3D-MSF.

The BM3D denoiser for 2D images operates by stacking self similar two dimensional patches along an additional dimension and collaboratively filtering the three dimensional group. It is an expansion of Non Local Means \cite{buades2011non}, however it denoises by enforcing sparsity in the transformed group.  The results of the filtered stack are then broadcast back to the original lower dimensional image using a weighted average in a scheme similar to Non Local Means. BM4D operates analogously for 3D volumetric images. 
Extending the analysis in \cite{hou2018image}, to search for block matched patches within a 3D search window, increasing the search space by a factor of $\alpha$ voxels on each side increases the computational complexity by $3\alpha^2$ for BM3D-MSF as opposed to $\alpha^3$ for PNP-BM4D.

\section{RESULTS}
In this section we compare the performance of our proposed method, BM3D-MSF,
against FBP, MBIR with an MRF (MBIR+MRF) prior as well as 
a PNP method with a BM4D prior (PNP-BM4D).
PNP-BM4D is formulated similarly to the system of updates in equations \ref{eq:pnp_x} through \ref{eq:pnp_u} with $s_2$ as a BM4D denoising operation on the volume $\hat{x}^{(k+1)}+u^{(k)}$.
The methods are evaluated on two sample CT scans of real patients \cite{exp_phantom}. 
The first instance is a 16 slice stack of $512\times 512$ pixel images of a real brain scan.
The second is a 24 slice stack of $512\times 512$ pixel images from a thoracic scan.
These density volumes are forward projected with a parallel beam CT system matrix 
using the svmbir \cite{svmbir-2024} python package. 
Measurement data is synthesized with 2000 views spanning $\pi$ radians,
512 $1\mbox{mm}\times 1\mbox{mm}$ detector channels with a $1\mbox{mm}$ spacing between slices, 
and uniform voxel size. 
This is considered the fully sampled measurement because it is the 
sampling density required to achieve less than $5\%$ Normalized Root Mean Squared Error 
(NRMSE) in the reconstruction made with the MBIR+MRF method that comes out 
of the box with svmbir.
The angular measurements are then subsampled by a factor 10
to simulate view sparsity, and sinogram slices are subsampled by a factor 
2 to simulate increasing pitch to produce two 100-view sinograms.
This represents a twenty-fold reduction in sampling density. 
Finally white Gaussian noise with standard deviation at $1\%$ of the respective measurement mean is added each sinogram.
The noisy sinogram data is first inverted using MBIR+MRF
and used as the initial condition for the two other PNP methods.
We use the scico python package \cite{scico-2024}
that has built-in support for implementing proximal map solvers for equation \ref{eq:pnp_z} using parallel beam CT forward model from svmbir as well as
integrating block matching denoisers into a PNP framework.

\begin{table}[]
\centering
\begin{tabular}{c|ccc|ccc|}
\cline{2-7}
                               & \multicolumn{3}{c|}{Brain Scan}                                                            & \multicolumn{3}{c|}{Thoracic Scan}                                                         \\ \cline{2-7} 
                               & \multicolumn{1}{c|}{NRMSE}          & \multicolumn{1}{c|}{PSNR (dB)}      & SSIM           & \multicolumn{1}{c|}{NRMSE}          & \multicolumn{1}{c|}{PSNR (dB)}      & SSIM           \\ \hline
\multicolumn{1}{|c|}{FBP}      & \multicolumn{1}{c|}{0.271}          & \multicolumn{1}{c|}{27.32}          & 0.420          & \multicolumn{1}{c|}{0.229}          & \multicolumn{1}{c|}{29.50}          & 0.562          \\ \hline
\multicolumn{1}{|c|}{MBIR+MRF} & \multicolumn{1}{c|}{0.109}          & \multicolumn{1}{c|}{35.22}          & 0.942          & \multicolumn{1}{c|}{0.097}          & \multicolumn{1}{c|}{36.99}          & 0.937          \\ \hline
\multicolumn{1}{|c|}{PNP+BM4D} & \multicolumn{1}{c|}{0.088}          & \multicolumn{1}{c|}{37.06}          & \textbf{0.972} & \multicolumn{1}{c|}{0.090}          & \multicolumn{1}{c|}{37.57}          & 0.947          \\ \hline
\multicolumn{1}{|c|}{BM3D-MSF} & \multicolumn{1}{c|}{\textbf{0.074}} & \multicolumn{1}{c|}{\textbf{38.56}} & 0.958          & \multicolumn{1}{c|}{\textbf{0.061}} & \multicolumn{1}{c|}{\textbf{41.05}} & \textbf{0.959} \\ \hline
\end{tabular}
\caption{Metrics for 24 slice volume reconstructions.}
\label{tab:my-table}
\end{table}

The fidelity between the ground truth volume $x^*$and its estimate $\hat{x}$ with $J$ voxel elements is quantified using NRMSE (eq (\ref{eq:nrmse})), Peak Signal to Noise Ratio (PSNR) (eq (\ref{eq:psnr})) and 
Structural Similarity (SSIM) \cite{wang2004image} with a window of side length seven.
Table \ref{tab:my-table} quantifies the performance of the different 
reconstruction techniques with the aforementioned metrics. 
Our propose method outperforms the others in all cases but one. 
  \begin{equation}
    \mbox{NRMSE}\left(\hat{x}, x^* \right) = \sqrt{\frac{\sum_j^J{\left([\hat{x}]_j - [x]_j^*\right)}^2}{\sum_j^J{[\hat{x}]_j^2}}}
    \label{eq:nrmse}
  \end{equation}\break
  \begin{equation}
    \mbox{PSNR}\left(\hat{x}, x^* \right) = 20 \times \frac{\log_{10}max_j\{[x^*]_j\}}{\sqrt{\sum_j^J{\left([\hat{x}]_j - [x^*]_j\right)}^{2} /J}}
    \label{eq:psnr}
  \end{equation}

For the PNP based methods, BM3D-MSF and PNP-BM4D, we track the primal and dual residuals \cite{scico-2024} defined in equations \ref{eq:primal} and \ref{eq:dual} and terminate when the residuals are at $5\%$ of there starting values. We observe empirically that  $\rho \approx 50$ produces primal and dual reduction at similar rates.
The measurement noise profile captured in $\Lambda$ is estimated 
using built in routines from svmbir. 
The denoising standard deviations for BM4D and the three BM3D denoisers are grid searched to achieve maximum PSNR scores. 

  \begin{equation}
    \mbox{primal} = \sqrt{\sum_{l=1}^L \left\{ \rho \left\lVert  \left( \hat{x}^{(k)} -
            \hat{z}_l^{(k)} \right) \right\rVert_2^2 \right\}}
    \label{eq:primal}
  \end{equation}\break
  \begin{equation}
    \mbox{dual} = \left\lVert \sum_{l=1}^L \rho \left( \hat{z}^{(k)}_l - \hat{z}^{(k-1)}_l \right) \right\rVert_2
    \label{eq:dual}
  \end{equation}

\begin{figure} [H]
\centering
\begin{tabular}{cccc}
\includegraphics[width=0.24\textwidth]{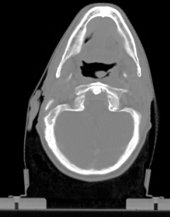} &
\includegraphics[width=0.24\textwidth]{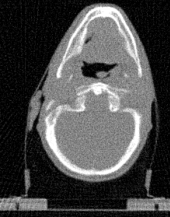} &
\includegraphics[width=0.24\textwidth]{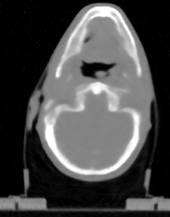} \\
\textbf{(a)}  & \textbf{(b)} & \textbf{(c)}  \\[6pt]
\end{tabular}
\begin{tabular}{cccc}
\includegraphics[width=0.24\textwidth]{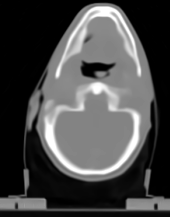} &
\includegraphics[width=0.24\textwidth]{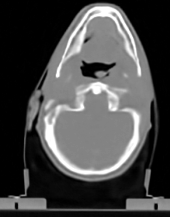} \\
\textbf{(d)}  & \textbf{(e)}  \\[6pt]
\end{tabular}
\caption{ \textbf{(a)} Ground truth image of brain scan. Reconstructions made at $20\times$ measurements reduction using
\textbf{(b)} FBP,
\textbf{(c)} Conventional MBIR+MRF,
\textbf{(d)} BM4D-PNP,
\textbf{(e)} BM3D-MSF}
\label{fig:brain_recon}
\end{figure}

\begin{figure} [H]
\centering
\begin{tabular}{ccc}
\includegraphics[width=0.40\textwidth]{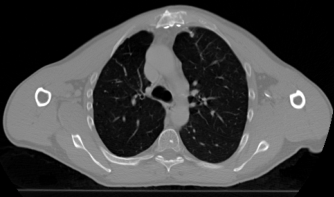} &
\includegraphics[width=0.40\textwidth]{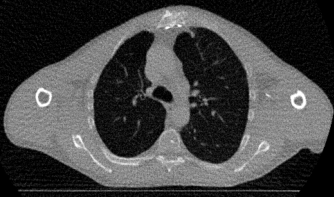} \\
\textbf{(a)}  & \textbf{(b)}  \\[6pt]
\end{tabular}
\begin{tabular}{ccc}
\includegraphics[width=0.40\textwidth]{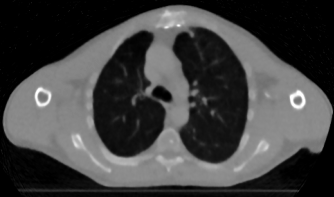} &
\includegraphics[width=0.40\textwidth]{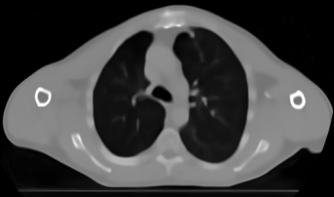} \\
\textbf{(d)}  & \textbf{(e)}  \\[6pt]
\end{tabular}
\begin{tabular}{cc}
\includegraphics[width=0.40\textwidth]{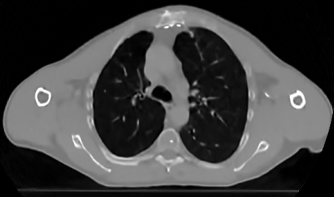} \\
\textbf{(d)}  \\[6pt]
\end{tabular}
\caption{ \textbf{(a)} Ground truth image of thoracic scan. Reconstructions made at $20\times$ measurements reduction using
\textbf{(b)} FBP,
\textbf{(c)} Conventional MBIR+MRF,
\textbf{(d)} BM4D-PNP,
\textbf{(e)} BM3D-MSF}
\label{fig:chest_recon}
\end{figure}

Figures \ref{fig:brain_recon} and \ref{fig:chest_recon} illustrate the improvement in reconstruction quality using denoiser based prior methods. 
In particular, our proposed method, BM3D-MSF is able to outperform the others and in both these metrics. 
In both test cases BM3D-MSF is able to preserve more fine-structural detail than PNP+BM4D, which is overly blurry.
The denoiser based priors completely suppress streak artefacts visible in the lower right corners of the brain reconstructions made using FBP and MBIR+MRF.
The BM3D-MSF prior is able to preserve more detail than PNP-BM4D in the patient's jaw line. 
In the thoracic reconstructions BM3D-MSF is able to remove streak artifacts present in FBP while retaining fine structures in the lung cavity and the sternum that are overly-smoothed in the other reconstructions. 

Intuitively, the advantage of our method over PNP+BM4D is likely to be of greater prominence at higher pitch values and additive noise. 
As the spatial variation in the image domain grows it becomes increasingly difficult to find similar 3D patches for block matching in BM4D compared to the much easier task of finding similar 2D patches for each of the individual BM3D operators. 
Additionally, the denoising variance of the BM3D operators along the $z$ axis can be individually tuned for varying pitch.

\section{CONCLUSIONS}
In this work, we demonstrate an effective strategy for the inversion of sparse
measurement CT for medical imaging applications using off the shelf denoisers 
as priors to build a fully 3D BM3D-MSF prior in a PNP framework.
It outperforms other denoisers in its class and in reconstruction quality. 
One drawback of MBIR methods including PNP is its computational time.
However, the modular nature of the MSF prior allows greater distributed implementation not only across the different equilibrium agents
but also across the axis orthogonal to the plane of each lower dimensional denoiser 
For reconstructions where the prior model computation dominates, 
a single update in this iterative framework only takes as long as the slowest operator.
Given sufficient computational resources for the prior model computation, this time is 
equivalent to one application of the lower dimensional denoiser to one slice in the stack. 
These run time gains using a distributed strategy is an important step in deploying this
innovation in the practical medical field.
The separate computation of the measurement fidelity model and the prior models also allows easy adaptation of the contributions herein not only to different CT scanner geometries such as helical and cone beam acquisition, but also to other imaging modalities such as MRI by simply updating the scanner geometry matrix.

\acknowledgments 
This project is primarily funded by DOE DRD seed funding program. This manuscript has been authored by ORNL, operated by UT-Battelle, LLC under Contract No. DE-AC05-00OR22725 with the U.S. Department of Energy.
This paper describes objective technical results and analysis. Any subjective views or opinions that might be expressed in the paper do not necessarily represent the views of the U.S. Department of Energy or the United States Government.


\bibliography{report} 
\bibliographystyle{spiebib} 

\end{document}